\documentclass[prd,twocolumn, superscriptaddress,floatfix,amsmath,amssymb,dvipsnames]
{revtex4-1}
\usepackage{amsmath,amsfonts,amsthm,amssymb,slashed,graphicx}
\usepackage{xcolor}
\usepackage{graphicx}
\usepackage{epstopdf}
\usepackage{array,booktabs}
\usepackage{hyperref}
\usepackage{bbm} 
\usepackage{mathtools} 
\usepackage{tabulary}
\usepackage{multirow}
\usepackage{array,arydshln}
\hypersetup{linktoc = all}  
\hypersetup{pdfborderstyle={/S/U/W 0.5}}
\usepackage{tikz}
\usepackage{tkz-tab}
\usepackage{caption}
\usepackage{subcaption}
\usetikzlibrary{shapes.geometric}
\usetikzlibrary{intersections}
\usetikzlibrary{calc}
\usetikzlibrary{decorations.pathmorphing}
\usepackage[toc]{appendix}
\newcounter{mycnt}
\newcommand{\be}{\begin{equation}}
\newcommand{\ee}{\end{equation}}
\newcommand{\ba}{\begin{eqnarray}}
\newcommand{\ea}{\end{eqnarray}}

\newcommand{\tcr}{\textcolor{red}}
\newcommand{\tcb}{\textcolor{blue}}

 \newcommand{\rbm}[1]{{\bf \tcr{[[[Robb: #1]]]}}}

\definecolor{babyblueeyes}{rgb}{0.63, 0.79, 0.95}
\definecolor{babyblue}{rgb}{0.54, 0.81, 0.94}
\definecolor{ballblue}{rgb}{0.13, 0.67, 0.8}
\definecolor{beaublue}{rgb}{0.74, 0.83, 0.9}
\definecolor{coolblack}{rgb}{0.0, 0.18, 0.39}
\definecolor{darkcyan}{rgb}{0.0, 0.55, 0.55}
\definecolor{darkmidnightblue}{rgb}{0.0, 0.2, 0.4}
\definecolor{darkgreen}{rgb}{0.0, 0.2, 0.13}
\definecolor{upforestgreen}{rgb}{0.0, 0.27, 0.13}
\definecolor{tropicalrainforest}{rgb}{0.0, 0.46, 0.37}
\DeclareCaptionJustification{justified}{\leftskip=0pt \rightskip=0pt \parfillskip=0pt plus 1fil}

\begin{document}

\title{Holographic Complexity and Charged Scalar Fields}

\author{Musema Sinamuli}
\email{cmusema@perimeterinstitute.ca}
\affiliation{Perimeter Institute for Theoretical Physics, Waterloo, Ontario, N2L 2Y5, Canada}

\author{Robert B. Mann}
\email{rbmann@uwaterloo.ca}
\affiliation{Department of Physics and Astronomy, University of Waterloo, Waterloo, Ontario N2L 3G1, Canada}
\affiliation{Perimeter Institute for Theoretical Physics, Waterloo, Ontario, N2L 2Y5, Canada}

\begin{abstract}

We construct a time-dependent expression  of the computational complexity of a quantum system which consists of two conformal complex scalar field theories in $d$ dimensions coupled to constant electric potentials and defined on the boundaries of a charged AdS black hole in $(d+1)$ dimensions. Using a suitable choice of the reference state, Hamiltonian gates and the metric on the manifold of unitaries, we find that the complexity grows linearly for a relatively large interval of time. We also remark that for scalar fields with very small charges the rate of variation of the complexity cannot exceed a maximum value known as the Lloyd bound. 

 \end{abstract}

\maketitle

\section{Introduction}

 Holographic dualities between quantum field and gravity theories continue to provide
us with interesting new information and approaches to understanding quantum gravity.
The AdS/CFT correspondence \cite{thelarge}, which is arguably the most reputed and explored, conjectures a duality between  a $d$-dimensional conformal field theory (CFT) formulated on a spacetime that can be considered to be the boundary of an asymptotically $(d+1)$-dimensional anti-de Sitter (AdS) spacetime.   A number of years ago a proposed  holographic connection between the entanglement entropy of a quantum (conformal) field theory and the Bekenstein-Hawking entropy of an AdS black hole arose \cite{holodev}.   

More recently
proposals have been put forward relating information theoretic quantities in   CFT
to geometric quantities in the corresponding bulk spacetime \cite{entangenough}.
One example \cite{gravitydual} was the proposal that a quantum information metric is dual to a codimension-1 time slice of an AdS black hole. Shortly afterward a quantum information quantity known as the computational complexity of the CFT was conjectured to be proportional to either a codimension-2 volume or to the action evaluated on a Wheeler-DeWitt (WDW) patch of the conformal diagram of an AdS black hole \cite{compucomp,complexityaction}, i.e.  
\begin{equation}\label{CIeq}
C=I_{\mbox{\tiny WDW}}/\pi.
\end{equation}

 Computational complexity can be thought of as the degree of difficulty to carry out a computational task.  In more precise and quantifiable terms, it has been defined as
 the minimum number of gates necessary to synthesize a unitary operator taking one
 state (called the reference) to another (called the target) \cite{ageometric}.   The minimum number of gates needed to approximate this unitary is geometrically interpreted as the minimum length (in the manifold of unitaries) between the identity operator and that unitary.
 
Proceeding from these conjectures, time-dependent expressions of  complexity were derived from investigations in the gravity side. These studies resulted in the observation that the rate of variation of the complexity at late time is proportional to twice the mass of the AdS black hole \cite{complexityaction,gravitational,complexity, onthetime}.  A proposed extension making use of a lattice for computing the complexity of free scalar field theory was then given \cite{circuitcomp}, from which  a time-dependent expression of the CFT complexity was formulated \cite{complexentang}. The complexity growth was found to be linear in a short interval of time of the order of the scrambling time of the system (for a system consisting of fast scramblers). Many attempts on the computation of the time dependence of the complexity were subsequently undertaken \cite{liouville,comparisonof,divergences1,evolution1,evolution2,timevol}  and led to similar conclusions.

A recent approach for computing complexity in continuous quantum many-body systems that exploited Gaussian states   \cite{towards} was subsequently generalized to 
make use of a different  choice of  gates  \cite{topological} to derive the time-dependent complexity in the CFT for a free scalar field theory, and this likewise led to similar results. 
However in this latter study it was shown that the  time interval could be much larger when considering a reference state (a thermofield double (TFD) state in this case) with a larger thermal circle.
   
Here we extend this previous study  \cite{topological} to charged AdS black holes and their dual CFTs, for which the conjectured complexity growth at late time is proportional to the difference of the mass and charge of the  black hole and does not exceed the Lloyd bound \cite{complexityaction,gravitational}. We shall  investigate a simple theory that consists of a free complex scalar field theory coupled to an electric potential.  We find that the complexity grows linearly for a larger interval of time since the scrambling time of the system is larger, and that for scalar fields with small charge the linear growth of complexity has the conjectured relationship to the   growth of the gravitational action evaluated on a WDW patch at late time.

 We organize our paper as follows: In section 2, we review the notion of   computational complexity in a CFT and emphasize   its geometrical formulation and interpretation.
 Section 3 is devoted to the exploration of  free complex scalar field theory in $d$-dimensions,  with an emphasis on the construction of the Hamiltonian of the theory as this plays a very central role in the computation of the complexity. Section 4 consists of the derivation of a time-dependent expression of the complexity following the Nielsen approach with a Hamiltonian gate and  a manifold of unitaries endowed with a Fubini-Study metric.
 In section 5, we provide a short discussion on our model based on an analogy with a system of N qubits, interacting in parallel in intervals of time \cite{bhmirrors}. We recapitulate our main results   and suggest further directions for future projects in section 6.

\section{Manifold and metric generation}

The main idea here is to introduce the notion of complexity using the geometrical approach employed by Nielsen \cite{ageometric}. Let us consider a unitary operator $U$ which maps a state referred to as the reference state to another state that we define as the target state. This unitary operator $U$ is thought of as an element of a manifold of unitaries, which is endowed with a local metric. The shortest distance between the identity operator $I$ and the unitary operator $U$ on the manifold of unitaries can be regarded as the minimum number of gates necessary to synthesize the unitary $U$ mapping the reference to the target state.
\vskip 5pt In the context of the present work we consider that the manifold of unitaries is considered to be the direct product of the noncompact group $SU(1,1)$ with the simplest unitary group $U_Q(1)$, i.e. $SU(1,1)\times U_Q(1)$.  The presence of the $U_Q(1)$ symmetry group is due to the fact that the CFT 
is supposed to carry an electric charge and is coupled to an electric potential as it is dual to a charged AdS black hole.  The above case is merely an extension of what has been studied in \cite{towards, topological} for a CFT dual to a Schwarzschild-AdS black hole whose manifold of unitaries is $SU(1,1)$.

Let us start with a unitary operator that belongs to the manifold $SU(1,1)\times U_Q(1)$, whose path parameter is $\sigma$, and which reads as   

\begin{eqnarray}
\label{unitary2}
U(\sigma)=e^{\int d^{d-1}k~g(\overrightarrow{k},\sigma)}
\end{eqnarray}
with
\begin{eqnarray}
g(\overrightarrow{k},\sigma)&=&\alpha_+(\overrightarrow{k},\sigma)K_+(\overrightarrow{k})+\alpha_-(\overrightarrow{k},\sigma)K_-(\overrightarrow{k})\nonumber\\
&+&\omega(\overrightarrow{k},\sigma)K_0(\overrightarrow{k})+\bar{\omega}(\overrightarrow{k},\sigma)\bar{K}_0(\overrightarrow{k}).
\end{eqnarray}
The functions $\alpha_\pm(\overrightarrow{k},\sigma), ~\omega(\overrightarrow{k},\sigma)$, and $\bar{\omega}(\overrightarrow{k},\sigma)$ are arbitrary functions of the CFT momentum $\overrightarrow{k}$ and the path parameter $\sigma$. $K_\pm, K_0$, and $\bar{K}_0$ are the generators of $SU(1,1)\times U_Q(1)$.
The directions that only give an overall phase to the state are modded out:
\begin{eqnarray}
\label{su11}
&&K_+=\frac{1}{2}c^\dagger_1c^\dagger_2\nonumber\\
&&K_-=\frac{1}{2}c_1c_2\nonumber\\
&&K_0=\frac{1}{4}(c^\dagger_1c_1
+c_2c^\dagger_2)\nonumber\\
&&\bar{K}_0=\frac{1}{4}(c^\dagger_1c_1
-c_2c^\dagger_2)
\end{eqnarray}
where $c_1=c_{\overrightarrow{k}},~c_2=\tilde{c}_{-\overrightarrow{k}}$   satisfy the commutation relations
\begin{eqnarray}
 {[c_{\small\overrightarrow{k}},c^\dagger_{\small \overrightarrow{k}^\prime}]}&=& {\delta^{d-1}({\small\overrightarrow{k}}-{\small\overrightarrow{k}^\prime})}\nonumber\\
 ~ {[\tilde{c}_{\small -\overrightarrow{k}},\tilde{c}^\dagger_{\small -\overrightarrow{k}^\prime}]}&=& {\delta^{d-1}({\small\overrightarrow{k}}-{\small\overrightarrow{k}^\prime})}
\end{eqnarray}
 and
\begin{eqnarray}
&&[K_+,K_-]=-K_0 ~~~~~~~[K_0,K_\pm]=\pm\frac{1}{2}K_\pm\nonumber\\
&&[\bar{K}_0,K_0]=0 ~~~~~~~[\bar{K}_0,K_\pm]=0 .
\end{eqnarray}

Since $\bar{K}_0$ commutes with the other generators it is straightforward to show that the unitary operator (\ref{unitary2}) can be expressed as  \cite{agroup}
\begin{eqnarray}
\label{unitary3}
U(\sigma)&=&e^{\int d^{d-1}k~\gamma_+(\overrightarrow{k},\sigma)K_+(\overrightarrow{k})}\nonumber\\
&\times &
e^{\int d^{d-1}k~\log(\gamma_0(\overrightarrow{k},\sigma))K_0(\overrightarrow{k})}\nonumber\\
&\times &e^{\int d^{d-1}k~\gamma_-(\overrightarrow{k},\sigma)K_-(\overrightarrow{k})}\nonumber\\
&\times &e^{\int d^{d-1}k~\tilde{\omega}(\overrightarrow{k},\sigma)\bar{K}_0(\overrightarrow{k})}
\end{eqnarray}
where the functions $\gamma_+(\overrightarrow{k},\sigma), \gamma_-(\overrightarrow{k},\sigma)$, and $\gamma_0(\overrightarrow{k},\sigma)$ read as 
\begin{eqnarray}
&&\gamma_{\pm}=\frac{2\alpha_{\pm}\sinh\Xi}{2\Xi\cosh\Xi-\omega\sinh\Xi}\nonumber\\
&&\gamma_0=(\cosh\Xi-\frac{\omega}{2\Xi}\sinh\Xi)^{-2}\nonumber\\
&&\Xi^2=\frac{\omega^2}{4}-\alpha_+\alpha_-.
\end{eqnarray}

To obtain the simplest possible form of (\ref{unitary3}) we impose the conditions \cite{towards} 
\begin{eqnarray}
&&K_-|R\rangle =0 
 \qquad  K_0|R\rangle =\frac{1}{4} \delta^{d-1} (0)|R\rangle\nonumber\\
&&\bar{K}_0|R\rangle =-\frac{1}{4} \delta^{d-1} (0)|R\rangle
\end{eqnarray}
on the reference state and that~ $\bar{\omega}^\ast(\overrightarrow{k},\sigma)=-\bar{\omega}(\overrightarrow{k},\sigma)$.

This last condition on the function $\bar{\omega}({\small\overrightarrow{k}},\sigma)$, in combination with the fact that the $U_Q(1)$ generator $\bar{K}_0$  commutes with the other generators, implies that the component along $\bar{K}_0$ will contribute just   an overall phase.  We shall see that the choice of these conditions shall  ease the computation by providing a suitable control function $\gamma_+$ when using particular Hamiltonian gates (which will be the case in the next sections).

These conditions lead to a target state of the form 
\begin{eqnarray}
|\Psi (\sigma)\rangle &=&N~e^{\int d^{d-1}k~ \gamma_+(\overrightarrow{k},\sigma)K_+(\overrightarrow{k})}|R\rangle\nonumber\\
\mbox{with}~~N&=& e^{\frac{1}{4}\delta^{d-1}(0)\int d^{d-1}k\log(\gamma_0(\overrightarrow{k},\sigma))}
\end{eqnarray}
in which only the factor involving $\gamma_+$ needs to be taken into account, with an overall phase  modded out.
From the normalization of that state it follows that $|\gamma_0|=1-|\gamma_+|^2$.
 
The reference state is chosen such that it is annihilated by the $c_{\overrightarrow{k}}$ and $\tilde{c}_{-\overrightarrow{k}}$
\begin{equation}
|R\rangle=|0,0\rangle
\end{equation}
and when omitting the variables and the integrals we find that
\begin{equation}
\label{zeromode}
|\Psi\rangle=Ne^{\gamma_+K_+}|0,0\rangle.
\end{equation}
The target state (\ref{zeromode}) becomes 
\begin{equation}
\label{zeromode1}
|\Psi\rangle=\sqrt{1-|\gamma_+|^2}\sum_{n}(\gamma_+)^n|n,n\rangle
\end{equation}
upon expanding into the basis of number (energy) state $|n\rangle ~(|E_n\rangle)$.
Inserting (\ref{zeromode1}) in the Fubini-Study metric,
\begin{equation}
\label{complexity}
ds^2_{FS}=\langle\delta\Psi|\delta\Psi\rangle-\langle\delta\Psi|\Psi\rangle\langle\Psi|\delta\Psi\rangle
\end{equation}
we obtain
\begin{equation}
\label{metric2}
ds_{FS}=\frac{|\delta\gamma_+|}{1-|\gamma_+|^2}.
\end{equation}
It appears that the above metric corresponds to the line element of the Poincaré disk whose associated manifold is the coset $SU(1,1)/U(1)$ of the manifold $SU(1,1)$.

\vskip 5pt The computational complexity is therefore defined as the shortest distance between two unitary transformations on the manifold of unitaries. Khaneja et al \cite{khaneja} have shown that finding the minimal length geodesic on the coset space $SU(2^n)/K$ (with $n\geq 1$ and $K$ a subgroup of $SU(2^n)$) of $SU(2^n)$ is equivalent to synthesize the unitary $U\in SU(2^n)$ in the minimum possible time.

In the case where the manifold of unitaries is $SU(1,1)$ with a coset $SU(1,1)/U(1)$ the complexity can be expressed as    
\begin{eqnarray}
\label{complexity1}
C^{(n)}&=&\min_{\gamma_+}\int^{s_f}_{s_i}d\sigma\sqrt[n]{\frac{V_{d-1}}{2}\int d^{d-1}k~|ds_{FS}(\sigma)/d\sigma|^n}\nonumber\\
&&
\end{eqnarray}
with ~$\gamma_+^\prime=\partial\gamma_+/\partial\sigma$ and $V_{d-1}$  the $(d-1)$-dimensional volume of a time slice.  $C^{(n)}$ is clearly an $L^{(n)}$ norm and for the sake of simplicity we will only focus on $(n=1)$ case whose complexity has the form    
\begin{eqnarray}
\label{complexity2}
C^{(1)}&=&\min_{\gamma_+}\int^{s_f}_{s_i}d\sigma ~\frac{V_{d-1}}{2}\int d^{d-1}k\frac{|\gamma_+^\prime|}{1-|\gamma_+|^2}. \nonumber\\
&&
\end{eqnarray}
The $C^{(1)}$ norm is obtained when gates for different momenta ($k$'s) are not allowed to act in parallel. 

\section{Complex scalar field}

The purpose of the current section is to provide enough technical background on the complex scalar field theory which will be very useful for what will follow in the next section. The complex scalar field will be defined in terms of the particle (antiparticle) creation and annihilation operators and so will the Hamiltonian and charge operators. Thus, the quantum gates built from the Hamiltonian and charge operators will also depend on these creation and annihilation operators. We shall also introduce the notion of pure Gaussian states for this particular theory. The ground state of the theory will be one of the Gaussian states as well as some other vacuum (for some momentum sector) obtained after Bogoliubov transformations of the creation and annihilation operators.

To this end, let us consider a complex scalar field theory in $d$ dimensions whose Hamiltonian is given by  
\begin{equation}
\label{hamiltonian0}
H_m=\frac{1}{2}\int d^{d-1}x~[\pi^\dagger\pi+\nabla\Phi^\dagger.\nabla\Phi+m^2\Phi^\dagger\Phi] 
\end{equation}
where $m$ is the mass of the field $\Phi (x)$, $\pi (x)$ is its conjugate momentum, and 
$\pi=\partial_0\Phi^\dagger,~\nabla\Phi\equiv\partial_i\Phi~~~(i=1,...,d-1)$. These functions obey the commutation rules
\begin{equation}
\label{commutation}
[\Phi (\overrightarrow{x}),\pi (\overrightarrow {x}^\prime)]=i\delta^{d-1}(\overrightarrow{x}-\overrightarrow{x}^\prime).
\end{equation}
In terms of the annihilation $a_k, ~b_k$ and creation operators $a^\dagger_k, ~b^\dagger_k$ of the particle and anti-particle respectively, the field and its associated momentum are explicitly given by 
\begin{eqnarray}
\label{scalar1}
\Phi (x)&=&\int d^{d-1}k\frac{1}{\sqrt{2\omega_k}}(a_k~e^{-ikx}+b^\dagger_k~e^{ikx})\nonumber\\
\pi (x)&=&-\int d^{d-1}k\frac{\sqrt{\omega_k}}{\sqrt{2}i}(a^\dagger_k~e^{ikx}-b_k~e^{-ikx})
\end{eqnarray}
with ~ $\omega_k=\sqrt{k^2+m^2}$.
 
Substituting   (\ref{scalar1}) into  (\ref{commutation}) we find
\begin{eqnarray}
&&[a_{\overrightarrow{k}},a^\dagger_{\overrightarrow{k}^\prime}]=\delta^{d-1}(\overrightarrow{k}-\overrightarrow{k}^\prime)\nonumber\\
&&[b_{\overrightarrow{k}},b^\dagger_{\overrightarrow{k}^\prime}]=\delta^{d-1}(\overrightarrow{k}-\overrightarrow{k}^\prime)
\end{eqnarray}
with all other commutators being zero. 

The Hamiltonian can be rewritten in a more useful form as a function of the particle (antiparticle) annihilation operators $a_k,~b_k$ and the particle (antiparticle) creation operators $a^\dagger_k,~b^\dagger_k$
\begin{equation}
\label{hamiltonian1}
H_m=\frac{1}{2}\int d^{d-1}k~\omega_k~ \big[a^\dagger_{\overrightarrow{k}}a_{\overrightarrow{k}}+
b^\dagger_{\overrightarrow{k}}b_{\overrightarrow{k}}+1\big]. 
\end{equation}
The charge operator associated with the field reads as 
\begin{equation}
Q=\frac{iq}{2}\int d^{d-1}x[\Phi^\dagger\dot{\Phi}-\Phi\dot{\Phi}^\dagger].
\end{equation}
As a function of the particle (antiparticle) annihilation and creation operators it becomes
\begin{equation}
\label{charge1a}
Q=\frac{q}{2}\int d^{d-1}k ~\big[a^\dagger_{\overrightarrow{k}}a_{\overrightarrow{k}}-
b^\dagger_{\overrightarrow{k}}b_{\overrightarrow{k}}\big].
\end{equation}
 The complex scalar field $\Phi$ and its Hermitian conjugate associated momentum  become 
\begin{eqnarray}
\label{scalarfield}
&&\Phi(\overrightarrow{k})=\frac{1}{\sqrt{2\omega_k}}(a_{\overrightarrow{k}}+b^\dagger_{-\overrightarrow{k}})\nonumber\\
&&\pi^\dagger(\overrightarrow{k})={\frac{\sqrt{\omega_k}}{\sqrt{2}i}}(a_{\overrightarrow{k}}-b^\dagger_{-\overrightarrow{k}})
\end{eqnarray}
in the momentum space. 
 In order to obtain a CFT we consider that the complex scalar field is massless   $(m=0)$.
 
 A pure Gaussian state $|S\rangle$ is a state  defined as \cite{towards}
\begin{equation}\label{GS}
\big[\sqrt{\frac{\alpha_k}{2}}\Phi(\overrightarrow{k})+\frac{i}{\sqrt{2\alpha_k}}\pi^\dagger(\overrightarrow{k})\big]|S\rangle =0
\end{equation}
where ~$\alpha_k=\omega_k$ corresponds to the ground state $|m\rangle$ ~of the theory. 

\begin{figure}
\centering
\begin{tikzpicture}[scale=1]
\draw(2,0)--(2,2)
node[midway, right, inner sep=1mm] {$\mbox{{\tiny CFT}}_2$};
\draw(0,2)--(0,0)
node[midway, left, inner sep=1mm] {$\mbox{{\tiny CFT}}_1$};
\draw[blue,thick](0,0)--(2,2);
\draw[blue](0,0)--(0.5,0.5)
node[midway, right, outer sep=1mm]
{$r_+$};
\draw[blue] (2,0)--(1.5,0.5)
node[midway, left, inner sep=1mm]
{$r_+$};
\draw[blue,thick] (2,0)--(0,2);
\draw[dashed] (0,2)--(0,4);
\draw[dashed] (2,2)--(2,4);
\draw[dashed] (0,0)--(0,-2);
\draw[dashed] (2,0)--(2,-2);
\draw[red,thick] (0,2)--(2,4);
\draw[red,thick] (2,2)--(0,4);
\draw[red,thick] (0,0)--(2,-2);
\draw[red,thick] (2,0)--(0,-2);
\draw[red] (0,-2)--(0.5,-1.5)
node[midway, right, outer sep=1mm]
{$r_-$};
\draw[red] (2,-2)--(1.5,-1.5)
node[midway, left, inner sep=1mm]
{$r_-$};
\draw[red] (0,2)--(0.5,2.5)
node[midway, right, outer sep=1mm]
{$r_-$};
\draw[red] (2,2)--(1.5,2.5)
node[midway, left, inner sep=1mm]
{$r_-$};
\end{tikzpicture}
\caption{The current diagram is the conformal diagram of a charged AdS black hole. The dashed and solid lines represent the singularity $r=0$ and boundaries $r=\infty$, respectively. The CFTs are defined on the boundaries (one CFT in each boundary). $r_+$ and $r_-$ are the outer and inner horizons, respectively. The blue and red lines are the horizons' radii. }\label{fig:M1}
\end{figure}
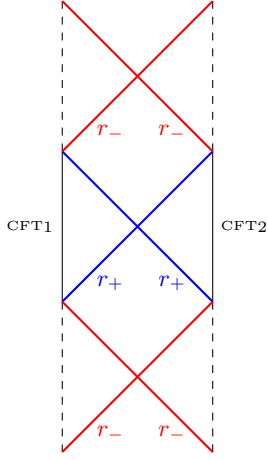

\section{Complexity in Conformal Field Theory}  

In this section we investigate  the time evolution of the computational complexity in conformal field theories in $d$ dimensions defined on the boundary of a charged AdS black hole. To start, we consider that the field theory consists  of  a free CFT coupled to an electric potential. In the case of an eternal black hole there exist two CFTs, one in each boundary of the conformal diagram of the black hole. We assume that one CFT consists in states of positive charge and is coupled to a positive electric potential $\mu$ and the other of negative charge states and coupled to a negative electric potential $-\mu$.

  The states in the CFTs dual to charged AdS black holes (see figure \ref{fig:M1}) are in the thermofield double (TFD) description with period $\beta$ and have the form \cite{complexityaction}   
\begin{eqnarray}
\label{tfd1}
|\mbox{TFD}_\mu(t)\rangle &\equiv & e^{-i(H_1+\mu Q_1)t_1} e^{-i(H_2-\mu Q_2)t_2}|\mbox{TFD}_\mu(0)\rangle \nonumber \\
&=& e^{-i[H_1+H_2+\mu(Q_1-Q_2)]t}|\mbox{TFD}_\mu (0)\rangle
\end{eqnarray}
 with $t_1=t_2\equiv t$ ~and~ $H_{1,2}$ the free Hamiltonians of fields 1 and 2 , respectively, and 
\begin{eqnarray}
\label{tfd0}
|\mbox{TFD}_\mu(0) \rangle &\equiv & N \sum_{n}e^{-\beta (E_n+\mu Q_n)/2}|E_n, Q_n\rangle_1|E_n,-Q_n\rangle_2\nonumber\\
&&
\end{eqnarray}
 with $N=\sqrt{1-e^{-\beta(\omega+\mu q)}}$. $|E_n, Q_n\rangle_1$, $|E_n, -Q_n\rangle_2$ are the eigenstates of the free Hamiltonians defined on the $\mbox{CFT}_{1,2}$ respectively. Considering that the system is made of harmonic oscillators, $E_n=n \omega$ and $Q_n= n q$~ are regarded as their corresponding energies and charges.

 In the context of a system of harmonic oscillators the state $|\mbox{TFD}_\mu(0)\rangle$ is expressed in the form \cite{eternalblack}
\begin{equation}
|\mbox{TFD}_\mu(0) \rangle= N e^{\int d^{d-1}k~ e^{-\beta(\omega_k+\mu q)/2}a^\dagger_{\overrightarrow{k}}
b^\dagger_{-\overrightarrow{k}}}|0\rangle
\end{equation}
 which  is annihilated by operators $c_{\overrightarrow{k}}$ and $\tilde{c}_{-\overrightarrow{k}}$ defined via  Bogoliubov transformations as
\begin{eqnarray}
\label{bogo}
c_{\overrightarrow{k}}&=&\cosh\theta_k a_{\overrightarrow{k}}-\sinh\theta_k b^\dagger_{-\overrightarrow{k}}\nonumber\\
\tilde{c}_{-\overrightarrow{k}}&=&\cosh\theta_k b_{-\overrightarrow{k}}-\sinh\theta_k a^\dagger_{\overrightarrow{k}}
\end{eqnarray}
with~ $\tanh\theta_k=e^{-\beta(\omega_k+\mu q)/2}$.  We see that despite the fact
that the   $U_Q(1)$ generator $\bar{K}_0$  commutes with the other generators, the parameter $\theta_k$ retains information about the charge.
 
 We consider that the CFT states in the boundaries are two-mode states whose one mode is on one side of the diagram (figure \ref{fig:M1}) and corresponds to states of a conformal complex scalar field theory with positive momentum $\overrightarrow{k}$ and the other mode on the other side of the diagram to a complex scalar field theory with negative momentum states $-\overrightarrow{k}$. The total Hamiltonian of the system according to (\ref{hamiltonian1}) and (\ref{charge1a}) reads     
\begin{eqnarray}
\label{totalham}
H&=&H_1+H_2+\mu (Q_1-Q_2)\nonumber\\
&=&\frac{1}{2}\int d^{d-1}k~\big[\omega_k[a^\dagger_1a_1+b^\dagger_2b_2+1
+a^\dagger_2a_2+b^\dagger_1b_1+1]\nonumber\\
&+& {\mu q [a^\dagger_1a_1+b^\dagger_2b_2-a^\dagger_2a_2-b^\dagger_1b_1]\big]}
\end{eqnarray}
where ~$\omega_k=k$, ~$a_1=a_{\overrightarrow{k}},~ ~b_1=b_{\overrightarrow{k}}, ~~a_2=a_{-\overrightarrow{k}}\mbox{and}~~b_2=b_{-\overrightarrow{k}}$.
The total Hamiltonian (\ref{totalham}) can be put into the following form 
\begin{eqnarray}
\label{totalham2}
H&=&{\frac{1}{2}\int d^{d-1}k~\big[(\omega_k+\mu q)(a^\dagger_1 a_1+b^\dagger_2 b_2+1)}\nonumber\\
&+& {(\omega_k-\mu q) (b^\dagger_1 b_1+a^\dagger_2 a_2+1)\big].}
\end{eqnarray}
Since the state $|\mbox{TFD}_\mu(0) \rangle$ is generated only by the operators $a_1$ and $b_2$, the second term in \eqref{totalham2} (which only has
operators involving $b_1$ and $a_2$) acts trivially on the reference state   and contributes only as an overall phase to the target state. We are therefore left with the operators involving $a_1$ and $b_2$, which will be the only ones taken into account in the unitary operator acting on the reference state. 
 
The total Hamiltonian (\ref{totalham2}) in the basis (\ref{su11}), using \eqref{bogo}, has operators of the form   
\begin{eqnarray}
a^\dagger_1a_1+b^\dagger_2b_2+1 &=& 2\cosh(2\theta_k)K_0+\sinh(2\theta_k)(K_++K_-)\nonumber\\
a^\dagger_1a_1-b^\dagger_2b_2-1 &=&4\bar{K}_0 ,
\end{eqnarray}
where we have included  $\bar{K}_0$ even though it does not appear in \eqref{totalham2}; this generator contributes  an overall phase factor to the target state, but in this particular case makes no phase contribution.

It follows that the resulting operator has components in the directions that correspond to the generators of $SU(1,1)\times U_Q(1)$.  
 
Therefore (\ref{tfd1}) becomes   
\begin{equation}
\label{tfd2}
|\mbox{TFD}_\mu(t)\rangle\equiv e^{\alpha_+ K_++\alpha_-K_-+\omega K_0}~|\mbox{TFD}_\mu(0)\rangle
\end{equation}
with  
\begin{eqnarray}
&&\alpha_{\pm}=-i~{(\omega_k+\mu q)}~t\sinh(2\theta_k)\nonumber\\
&&\omega=-2i~{(\omega_k+\mu q)}~t\cosh(2\theta_k).
\end{eqnarray}
Equation (\ref{tfd2}) read as
\begin{equation}
\label{tfd4}
|\mbox{TFD}_\mu (t)\rangle\equiv e^{\gamma_+K_+}e^{\log(\gamma_0)K_0} e^{\gamma_-K_-}~|\mbox{TFD}_\mu(0)\rangle 
\end{equation}   
when using the transformation of the unitary operator (\ref{unitary3}).

\vskip 5pt The above state \eqref{tfd4} is reduced to (\ref{zeromode}) and (\ref{zeromode1}) when following the same steps,  
with the control functions
\begin{eqnarray}
&&\gamma_{\pm}=\frac{-i\sinh(2\theta_k)\sin\Xi}{\cos\Xi+i\cosh(2\theta_k)\sin\Xi}\nonumber\\
&&\Xi={(\omega_k+\mu q)}~t~~~~\mbox{and}~~~~\omega_k=k.
\end{eqnarray}
 
  In terms of the path parameter $\sigma$ the control function $\gamma_+$ is written as
\begin{eqnarray}
&&\gamma_\pm(k,\sigma)=\frac{-i\sinh(2\theta_k)\sin\Xi}{\cos\Xi+i\cosh(2\theta_k)\sin\Xi}\nonumber\\
&&\Xi={(k+\mu q)}t~\sigma.
\end{eqnarray}
 The control function $\gamma_+=\gamma_+(k,\sigma)$, as a function of $\sigma$, verifies the conditions
\begin{eqnarray}
\gamma_+(k,s_i)&=&0~~~~\mbox{and}\nonumber\\
\gamma_+(k,s_f)&=&\frac{-i\sinh(2\theta_k)\sin ({(k+\mu q)}t)}{\cos ({(k+\mu q)}t)+i\cosh(2\theta_k)\sin ({(k+\mu q)}t)}\nonumber\\
&&
\end{eqnarray}
which correspond to the reference and target state respectively. The time-dependent control function $\gamma_+$ will obviously imply a time-dependent complexity.

\vskip 5pt Inserting the control function $\gamma_+$ into the complexity (\ref{complexity2}) we get a time-dependent expression of the form   
\begin{eqnarray}
C^{(1)}(t)&=&\min_{\gamma_+}\int^{s_f}_{s_i}d\sigma ~\frac{V_{d-1}}{2}\int d^{d-1}k~\frac{|\gamma_+^\prime|}{1-|\gamma_+|^2}\nonumber\\
&=&{V_{d-1}\Omega_{\kappa,d-2} \bigg[2^{d-1}\beta^{-d}\Gamma (d)} \nonumber \\
&&{\times\left( \textrm{Li}_d(e^{-\mu q/2}) -  \textrm{Li}_d(- e^{-\mu q/2})\right)}\nonumber\\
&+& {\mu q~2^{d-2}\beta^{-(d-1)}\Gamma (d-1)} \nonumber \\ 
&&{\times\left( \textrm{Li}_{d-1}(e^{-\mu q/2}) -  \textrm{Li}_{d-1}(- e^{-\mu q/2})\right)\bigg]~t}\nonumber\\
&&
\label{complexC1}
\end{eqnarray}
as detailed in eq. \eqref{appendix1} in the appendix, where $ \textrm{Li}_d (z)$ is the polylog function.  We see from this expression that the complexity evolves linearly in time.

For $q$ very small eq. \eqref{complexC1} reads as
\begin{eqnarray}
\label{complexC2}
C^{(1)}(t)&=&{V_{d-1}\Omega_{\kappa, d-2} \big[(2^d-1)\beta^{-d}\Gamma (d)\zeta(d)}\nonumber\\
&-&{(d-2)(2^{d-1}-1)\beta^{-(d-1)}\Gamma (d-1)\zeta (d-1)\mu q\big]~t}.\nonumber\\
&&
\end{eqnarray}  

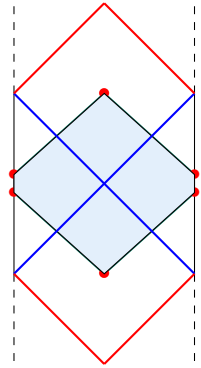
\begin{figure}
\centering
\begin{tikzpicture}[scale=1.2]
\draw(2,0)--(2,2);
\draw(0,2)--(0,0);
\draw[thick,tropicalrainforest](1,2)--(0,1.1);
\draw[thick,tropicalrainforest](1,2)--(2,1.1);
\draw[thick,tropicalrainforest](0,0.9)--(1,0);
\draw[thick,tropicalrainforest](1,0)--(2,0.9);
\node [red] at (1,2) {\textbullet};
\node [red] at (0,1.1) {\textbullet};
\node [red] at (1,2) {\textbullet};
\node [red] at (2,1.1) {\textbullet};
\node [red] at (0,0.9) {\textbullet};
\node [red] at (1,0) {\textbullet};
\node [red] at (1,0) {\textbullet};
\node [red] at (2,0.9) {\textbullet};
\draw[fill=babyblueeyes!30](1,2)--(0,1.1)--(0,0.9)--(1,0)--(1,0)--(2,0.9)--(2,1.1)--(1,2)--cycle;
\draw[dashed] (0,0)--(0,-1);
\draw[dashed] (2,0)--(2,-1);
\draw[dashed] (0,2)--(0,3);
\draw[dashed] (2,2)--(2,3);
\draw[blue,thick] (0,0)--(2,2);
\draw[blue,thick] (0,2)--(2,0);
\draw[red,thick] (2,2)--(1,3);
\draw[red,thick] (0,2)--(1,3);
\draw[red,thick] (0,0)--(1,-1);
\draw[red,thick] (2,0)--(1,-1);
\end{tikzpicture}
\caption{In this diagram we depict the WDW patch, on which is evaluated the action. The light blue area is the bulk, the green lines are the null geodesics and the red dots are the joints (intersections of null-null  or null-timelike geodesics). The different contributions to the action come from the bulk, the surface terms and the joint terms. The late time variation of the action evaluated on the WDW is proportional to the mass and the charge of the charged AdS black hole and is less or equal to a quantity known as the Lloyd bound. }\label{fig:M2}
\end{figure}

In order to understand the contribution of the second term in \eqref{complexC2}, we define the total energy of the neutral scalar field (q=0) as (see \eqref{scalarenergy} and \eqref{charge1} in the appendix)
\begin{eqnarray}
\label{energy1}
E&=&V_{d-1}\int d^{d-1}k~ \omega_k \frac{e^{-\beta \omega_k}}{1-e^{-\beta \omega_k}}\nonumber\\
&=&V_{d-1}\Omega_{\kappa, d-2}\beta^{-d}\Gamma (d)\zeta (d)
\end{eqnarray}
and its total charge (when $q$ is very small)
\begin{eqnarray}
\label{charge0}
Q&=&V_{d-1}\int d^{d-1}k~ q~ \frac{e^{-\beta (\omega_k +\mu q)}}{1-e^{-\beta (\omega_k +\mu q)}}\nonumber\\
&=& q ~V_{d-1}\Omega_{\kappa, d-2}\beta^{-(d-1)}\Gamma (d-1)\zeta (d-1).
\end{eqnarray}  
Hence the complexity \eqref{complexC1} has the form
\begin{eqnarray}
\label{adscomplex}
C^{(1)}(t)&=&{[(2^d-1)E-(d-2)(2^{d-1}-1)\mu Q]~ t}\nonumber\\
&=&{[n_d~E- (d-2)n_{d-1}~\mu Q]~t}
\end{eqnarray}
with $n_d=(2^d-1)$ a dimensionless constant.
\vskip 5pt The rate of change of the complexity for very large $t$ is
\begin{eqnarray}
\label{rated2}
{\frac{dC(t)}{dt}}={n_d E - (d-2)n_{d-1}\mu Q.}
\end{eqnarray}
Equation (\ref{rated2}) implies that the variation of the complexity with respect to time at late time is proportional to the  total energy $E$ of the neutral scalar field and the total charge $Q$ ($q$ very small) of the complex scalar field theory.  When the energy of the neutral scalar field is identified with the mass of the AdS black hole and the charge of the complex scalar field with the charge of the black hole we find that the complexity is proportional to the action evaluated on the WDW patch (see figure \ref{fig:M2}).

We pause to comment on the correspondence between a CFT that consists of a charged scalar field theory coupled to an electric potential, and a charged AdS black hole.
Charged AdS black holes are solutions of Einstein-Maxwell (EM) truncation of gauged supergravities. Einstein-Maxwell-AdS (EM-AdS) truncations are associated with rotating branes (particularly the $\mbox{EM-AdS}_4$ and $\mbox{EM-AdS}_5$), and dual field theories are thought to arise on the world volume of these branes \cite{chargedadsbh}. 
Although the $\mbox{EM-AdS}_7$ is not related to a rotating-brane truncation of the $\mbox{AdS}_7\times \mbox{S}^4$ gauge supergravity (therefore its dual field theory cannot be declared to live on a rotating $\mbox{M}5$-brane world-volume), AdS holography can still be thought of as a phenomenon that exists independently of string and M-theory contexts and dual field theories are expected beyond $d=4,5$ \cite{adsspace0,gravinstanton} .

 According to \cite{chargedrot} a $\mbox{CFT}_d$ dual to a charged $\mbox{AdS}_{d+1}$ black hole corresponds to a theory in an Einstein universe with a chemical potential. This statement can be explained by the fact that an $\mbox{AdS}_{n+1} \times M^m$ spacetime is dual to a $\mbox{CFT}_n$ defined in a space with the topology of the $\mbox{AdS}_{n+1}$ and that the isometries of the manifold $M^m$ imply global symmetries of the boundary CFT \cite{chargedadsbh}. This can also be extended to gauge symmetries ($\mbox{SU(N)}$ is our case, with $\mbox{U(1)}$   a subgroup thereof).
Since the thermal properties of EM-AdS black holes are consistent with field theory interpretations \cite{chargedadsbh} , we infer that the $\mbox{U(1)}$ charge can be regarded as a thermodynamic quantity for both the charged AdS and its dual CFT. However, it is still necessary to clarify what the CFT consists of.

 As an illustration, let us consider a
$D=4,~ {\cal{N}}=4$ super Yang-Mills theory dual to spinning branes ($10d$ IIB gauged supergravity on $\mbox{AdS}_5\times \mbox{S}^5$) whose $\mbox{EM-AdS}_5$ arise from the truncation. This theory contains six real scalar fields $X^i~(i=1, ..., 6)$. For the sake of simplicity in the complexity computation the free scalar field Hamiltonian can be truncated from the total Hamiltonian of the system and treated as a theory in its right (though it would be more consistent to deal with the whole CFT) . Thus we advocate that there may exist  a $d$- dimensional theory of scalar fields that is part of a larger CFT in $d$ dimensions dual to a charged $(d+1)-$dimensional AdS black hole.

The variation of the action at late time is bounded by the Lloyd bound \cite{complexityaction,gravitational,complexity}, i.e, 
\begin{equation}
\frac{dI_{\mbox{\tiny WDW}}}{dt}\bigg|_{t\rightarrow\infty}
\leq 2(M-\mu Q)
\end{equation}
though we note a recent claim that for anisotropic black branes this bound can be violated
\cite{HosseiniMansoori:2018gdu}.

\section{Discussion}

This section covers a short discussion on an deductive study based on the system of $N$ qubits investigated in \cite{bhmirrors} by Hayden and Preskill as it yields a way to assess the scrambling time of our system. 

 They studied a system which is thought of as a parallel processing model, i.e. a system where multiple disjoint pairs of qubits are allowed to interact simultaneously, considering $N$ qubits where every qubit interacts once in each time step ($\beta$ is the time between the steps). From these considerations it follows that the total number of $U(4)$ operators required to scramble the system is $N\log N$ and the minimum scrambling time is $t_\ast=\beta\log N$.

 Although the system described above is not a conventional Hamiltonian system (since it consists of repeated discrete random unitary operations in parallel), we can still regard  it as a discrete model, whose   discrete step time is identified with a time interval of the order of the inverse of the energy per degree of freedom \cite{fastscamb}. Indeed, this time interval is the one during which each degree of freedom interacts once. According to  \eqref{energy1} this interval of time scales as $\Delta t\sim\beta$ for our Hamiltonian system.

For systems referred to as fast scramblers $\beta$ is very small. In the case where $\beta$ is large (which corresponds to a large scrambling time) the time variation of the complexity \eqref{rated2}  is
\begin{equation}
\frac{dC}{dt}\bigg|_{t\rightarrow\beta} \propto n_d \zeta (d) \beta^{-d}
- n_{d-1}\zeta (d-1)\mu q \beta^{-(d-1)}
\end{equation} 
using \eqref{energy1} and \eqref{charge0}.   
These correspond to charged AdS black holes with small mass and charge.
Moreover, it has been posited that thermal properties of AdS black holes can be reinterpreted as those of their corresponding CFTs at the same temperature \cite{adsspace} . From these properties we infer that the mass of a Schwarzschild-AdS black hole is proportional to the inverse of the dth power of $\beta$ (see eq. \eqref{adsmass1}). 

In the limit $\beta\rightarrow\infty$, the reference state $|\mbox{TFD}_\mu(0)\rangle$ becomes the ground state $|0\rangle$ of the theory. The energy E and the charge Q vanish, and so does the rate of variation of the complexity.

\section{Conclusion}

 We derived a time-dependent expression of the computational complexity for a $d$-dimensional CFT, which consists of a complex scalar field theory coupled to a constant electric potential, defined in the boundary of a charged asymptotically AdS black hole in $d+1$ dimensions. We observed that the complexity grows linearly for a large interval of time 
 since the scrambling time of the system is large. This can be explained by considering a reference state, i.e., the TFD at initial time, with a large thermal circle ($\beta$ large). While $\beta$ is very small our results conform with those of the cMERA circuit \cite{entanglementrenorm,holographicgeom1,holographicgeom2}, in which the complexity grows as $V_{d-1}\beta^{-(d-1)}$ in a short interval of time proportional to $\beta$.

For complex scalar fields with very small charge ($q\rightarrow 0$) the linear growth of the complexity ($\beta$ large) can be compared (up to some constant $n_d$) to the growth of the gravitational action evaluated on a WDW patch at late time. The latter is bound by a limit referred to as the Lloyd bound.   
 
  For future directions it would be of interest to investigate in the time dependence of the complexity of a theory in which a complex scalar field is coupled to a variable electric potential ($\mu=\mu(x)$ or a local gauge). Furthermore, theories involving fermionic and gauge fields shall constitute good candidates to dig into for the study of time-dependent complexities of CFTs dual to charged AdS black holes.
   
\section*{Appendix}
\subsection{Complexity evaluation}
\renewcommand{\theequation}{A-\arabic{equation}}
\setcounter{equation}{0}

In this subsection we perform explicit calculations to derive the final form of the computational complexity $C^{(1)}(t)$. The complexity as defined in the previous sections becomes (for $q$ very small)
\begin{eqnarray}
\label{appendix1}
C^{(1)}(t)&=&\min_{\gamma_+}\int^{s_f}_{s_i}d\sigma~\frac{V_{d-1}}{2}\int d^{d-1}k~\frac{|\gamma_+^\prime|}{1-|\gamma_+|^2}\nonumber\\
&=&\int^{s_f}_{s_i}d\sigma~\frac{V_{d-1}}{2}\int d^{d-1}k~|{(\omega_k+\mu q)} t\sinh(2\theta_k)|\nonumber\\
&=&V_{d-1}~t~\Omega_{\kappa, d-2}\int {(k^{d-1}+\mu q~ k^{d-2})}\nonumber\\
&&\quad\quad\quad\quad\quad\quad\quad
\times{\frac{e^{-\beta (k+\mu q)/2}}{1-e^{-\beta (k+\mu q)}}dk}\nonumber\\
&=&{V_{d-1}\Omega_{\kappa,d-2}\bigg[\beta^{-d}\Gamma (d)\sum_{n=0}\frac{e^{-(n+1/2)\beta\mu q}}{(n+1/2)^d}} \nonumber\\
&+&{\mu q\beta^{-(d-1)}\Gamma (d-1)\sum_{n=0}\frac{e^{-(n+1/2)\beta\mu q}}{(n+1/2)^{d-1}}\bigg]~t} \nonumber\\
&=&  {V_{d-1}\Omega_{\kappa,d-2} \bigg[2^{d-1}\beta^{-d}\Gamma (d)} \nonumber \\
&&{\times\left( \textrm{Li}_d(e^{-\mu q/2}) -  \textrm{Li}_d(- e^{-\mu q/2})\right)}\nonumber\\
&+& {\mu q~2^{d-2}\beta^{-(d-1)}\Gamma (d-1)} \nonumber \\ 
&&{\times\left( \textrm{Li}_{d-1}(e^{-\mu q/2}) -  \textrm{Li}_{d-1}(- e^{-\mu q/2})\right)\bigg]~t}\nonumber\\
&=&{V_{d-1}\Omega_{\kappa, d-2} \big[(2^d-1)\beta^{-d}\Gamma (d)\zeta(d)}\nonumber\\
&-&{(d-2)(2^{d-1}-1)\beta^{-(d-1)}\Gamma (d-1)\zeta (d-1)\mu q\big]~t}\nonumber\\
&&
\end{eqnarray}
up to the leading order in $q$, with the control function 
\begin{equation}
\label{appendix2}
\gamma_+=\frac{-i\sinh(2\theta_k)\sin ({(k+\mu q)}t\sigma)}{\cos ({(k+\mu q)}t\sigma)+i\cosh(2\theta_k)\sin ({(k+\mu q)}t\sigma)}
\end{equation}
yielding the final expression
\begin{equation}
\label{appendix3}
\frac{|\gamma_+^\prime|}{1-|\gamma_+|^2}={(\omega_k+\mu q)} t\sinh(2\theta_k)
\end{equation}
where 
\begin{equation}
\sinh(2\theta_k)
=\frac{2e^{-\beta(\omega_k+\mu q)/2}}{1-e^{-\beta(\omega_k+\mu q)}}.
\end{equation}  
\subsection{Total energy of the scalar field}
\renewcommand{\theequation}{B-\arabic{equation}}
\setcounter{equation}{0} 

 The current subsection is devoted to the computation of the total energy of the (neutral) complex scalar field and the total charge of the complex scalar field (for $q$ very small) knowing the probability densities of the Hamiltonian eigenstates $|E_n,Q_n\rangle_1|E_n,-Q_n\rangle_2$ (we will rather use the simplified notation $|n,n\rangle$ ). 
 
\vskip 5pt Considering the $|\mbox{TFD}_\mu(0)\rangle$ state in \eqref{tfd0} the density matrix is obtained from the expression
\begin{eqnarray}
\rho&=&\mbox{Tr}(|\mbox{TFD}_\mu(0)\rangle\langle \mbox{TFD}_\mu(0)|)\nonumber\\
&=&(1-e^{-\beta (\omega_k +\mu q)})\sum_{n_k} e^{-\beta n_k (\omega_k +\mu q)}|n_k\rangle\langle n_k|\nonumber\\
&&
\end{eqnarray}
after tracing over the states $|n_k\rangle_2$.
 We find that the probability densities of the eigenstates are
\begin{equation}
(1-e^{-\beta(\omega_k +\mu q)})e^{-\beta n_k(\omega_k +\mu q)}.
\end{equation}  
From these densities we obtain that the total energy of the neutral scalar field ($q=0$) reads as
\begin{eqnarray}
E&=&\sum_{n_k}n_k\omega_k e^{-\beta n_k\omega_k}(1-e^{-\beta\omega_k})\nonumber\\
&=&\omega_k\frac{e^{-\beta\omega_k}}{1-e^{-\beta\omega_k}}.
\end{eqnarray}
Restoring the integrals we get
\begin{eqnarray}
\label{scalarenergy}
E&=&V_{d-1}\int d^{d-1}k~ \omega_k~ \frac{e^{-\beta\omega_k}}{1-e^{-\beta\omega_k}}\nonumber\\
&=&V_{d-1}\int d^{d-1}k~ k~ \frac{e^{-\beta k}}{1-e^{-\beta k}}\nonumber\\
&=&V_{d-1}\Omega_{\kappa,d-2}\beta^{-d}\Gamma(d)\zeta (d).
\end{eqnarray}
The total charge of the complex scaler field (for $q$ very small) is 
\begin{eqnarray}
Q&=&q\sum_{n_k}n_k e^{-\beta n_k(\omega_k+\mu q)}(1-e^{-\beta(\omega_k+\mu q)})\nonumber\\
&=&q~\frac{e^{-\beta (\omega_k+\mu q)}}{1-e^{-\beta(\omega_k+\mu q)}}
\end{eqnarray}
and after restoring the integrals the above expression reads
\begin{eqnarray}
\label{charge1}
Q&=&q~V_{d-1}\int d^{d-1}k~\frac{e^{-\beta(\omega_k+\mu q)}}{1-e^{-\beta(\omega_k+\mu q)}}\nonumber\\
&=&q~V_{d-1}\Omega_{\kappa, d-2}\beta^{-(d-1)}\Gamma(d-1)\zeta(d-1).
\end{eqnarray}
\subsection{Mass of the Schwarzschild-AdS black hole}
\renewcommand{\theequation}{C-\arabic{equation}}
\setcounter{equation}{0}

 The present subsection is intended to derive the mass of a Schwarzschild-AdS black hole as well as to show its dependence on the period $\beta$ of the thermal circle of the TFD state. For the sake of simplicity we will only restrict our interest to the planar case ($\kappa=0$).
 
 In the coordinate $z=l/r$ a planar Schwarzschild-AdS black hole in $d+1$ dimensions admits the metric
\begin{eqnarray}
\label{Scharzschild-AdS}
ds^2&=&\frac{l^2}{z^2}[-h~dt^2+dz^2/h+d\Sigma^2_{\kappa,d-1}]\nonumber\\
h&=&1-(z/z_0)^d
\end{eqnarray}
where ~$z_0^d=l^{d-2}/\omega^{d-2}$~ and ~$\omega^{d-2}=r_+^d/l^2$. ~$r_+$ is the horizon radius in radial coordinates and ~$l$~ is the AdS radius. The mass  of this black hole reads as 
\begin{equation}
\label{massterm}
M=\frac{d-1}{16\pi G_N}~\Omega_{0,d-1}~\omega^{d-2}
\end{equation}
and its temperature
\begin{equation}
\label{adstemperature}
T=\frac{d}{4\pi z_0 l}.
\end{equation}
 Since the thermal properties of the AdS black hole can be regarded as those of the dual CFT \cite{adsspace} whose   temperature is the inverse of the period $\beta$ it follows, when substituting $z_0\sim  \beta$ from \eqref{adstemperature} and \eqref{Scharzschild-AdS} in \eqref{massterm}, that
\begin{equation}
\label{adsmass1}
M\sim\beta^{-d}.
\end{equation}
 
\section*{Acknowledgments} 
This work was supported in part by the Natural Sciences and Engineering Research Council of Canada.

\end{document}